# ON ESTIMATING THE RETURN VALUES
# FOR WIND SPEED AND WIND-WAVE HEIGHTS

Polnikov V.G., Gomorev I.A.

A.M. Obukhov Institute of Atmospheric Physics

Russian Academy of Sciences, Moscow, 119017 Russia,

e-mail: **polnikov@mail.ru**

Abstract

To estimate the return values for wind and wind-waves time series, we propose to use the extrapolation of a polynomial approximation built for a small part of the tail of provision function estimated for the series considered. A possibility of optimizing an accuracy of the polynomial approximation is stated, which is realized by varying both the length of the proper part for provision function and the exponent of the polynomial. The quality criteria for constructing approximation are proposed, allowing to increase the approximation accuracy. On example of the wind-reanalysis data and the results of numerical modeling of wind-wave heights in the Indian Ocean for the period 1980-2010, the proposed method was applied to obtain estimates of the return values at a number of fixed points in the ocean. It is shown that the method of polynomial approximation has a distinct advantage in estimating the return values with respect to the well-known method of initial Weibull distribution, taken for comparison.

# 1 Introduction

In a generalized sense, the values having a certain level of statistical provision are called as the return values for a random series considered [2]. Usually they are found on a basis of processing a random series of some geophysical magnitude (wind speed, wave height, current speed, temperature, pressure, etc.). The return values, which can be observed (to appear) at a given point once for the period of 30, 50, or 100 years, are the most popular in practice. When return values are considered for the period extending 30 years, they are called as climatic ones.

All kinds of return values of wind and waves are widely used in the risk assessment of marine activity: coastal construction, safety of navigation, marine oil and gas industry, environmental wasting, resort business, etc. For this purpose, a large number of methods are derived [2], basing on which numerous handbooks, atlases, and manuals are prepared, representing return values for wind and wind-wave fields in a wide variety of areas (for example, [1, 5]). The international scientific community has developed a general concept of providing the proper wind and wave information necessary for exploration resources of the oceans and seas [5, 7]. Despite of further discussing the return values estimates on the example of series for wind speed and wind-wave heights, our considerations are applied to any random geophysical series.

The most widely spread methods of return values assessment, described in detail in literature (for example, [2]), are as follows: the initial distribution method (IDM); the method of annual maxima (AMS - Annual Maxima Series); the method of overcoming the threshold (POT - Peak Over Threshold), and the method of percentile functions (BOULVAR).

All these methods are based on executing the following three procedures:

1) Calculation the histogram, $H(W_i)$, for considered time series of random values W, with a discrete of constructing $\Delta W$ (here $W_i$ means the interval $[W_i < W < W_i + \Delta W]$);

2) Evaluating the probability provision function, $F(W)$, on the basis of histogram $H(W_i)$, performed with the formula

$$F(W) = 1 - \int_0^W P(W)dW \approx 1 - \sum_{W_i=0}^{W_i=W} H(W_i) = \sum_{W_i=W}^{W_i=W_M} H(W_i), \qquad (1)$$

where $P(W)$ is the probability density function of the random variable, and $W_M$ is the maximum value of the time series considered. Here it is useful to we note that $F(W_M) = H(W_M)$;

3) Extrapolating provision functions $F(W)$ beyond the existing maximum value $W_M$, by means of constructing the analytical approximation, $F_{ap}(W)$, of function $F(W)$.

The main problem in evaluating return values consists in fulfillment the requirements of uniqueness and authenticity for performing extrapolation $F_{ar}(W)$ beyond the maximum value of the time series considered. It is clear that any uncertainty in the extrapolation leads to inaccuracies in obtaining return values estimates. Therefore, the problem of constructing an extrapolation provision function turns often out into a separate study (see description of the methods in [2, 6]).

Typically, solving the return values problem needs justification of the statistics choice, defining the extrema of the quantity considered. Thus, traditionally [2,6,7] it comes out of mind that statistics of natural processes may vary due to different dynamics of geophysical processes on different scales [3]. Therefore, the probability distribution, describing statistics of some geophysical quantity $W$ on different scales of its variability, may not exist. As a result, there is no evidently priority ways of solving the problem of extrapolating provision functions $F(W)$ at now. Furthermore, the question of approximation accuracy is generally not discussed in presently known approaches to this problem (see, [2,6,7]).

Therefore, it is needed to search for methods of constructing approximations with flexible statistics, applicable to empirically derived provision functions $F(W)$, and allowing to control an accuracy and confidence of approximation. One of such method is proposed in this paper.

### 2 Statement of the problem

From methodological point of view, the problem of constructing an analytical approximation of provision function has the following main questions: 1) what domain of function $F(W)$ should be used; 2) what statistical distribution should be attracted; 3) what requirements must be met to ensure reliability of the return values estimates.

Let us explain the essence of these issues on the example of the ERA-Interim wind-reanalysis series for the period 1980-2010yy, having discrete of 6 hours, and wind-wave height 3-hours series calculated from these wind data in the Indian Ocean with the modified model WAM4 [4, 5].

First of all, we consider the point of selecting the domain of provision function $F(W)$, used for building its approximation. To do this, let us consider the shapes of histograms $H(W_i)$ and functions $F(W)$ defined by the formula (1). It is known that the histogram may be both simple shaped (unimodal) and complex shaped (multimodal). Herewith, the appropriate provision functions $F(W)$, being the subject of the approximation, always exhibit a smooth behavior (Fig. 1a). Therefore, to construct an analytical approximation $F_{ap}(W)$, intended for its extrapolation beyond the observed maximum value $W_M$, one should be restricted by a rather small domain of the function $F(W)$, lying upper the latest mode of the histogram considered.

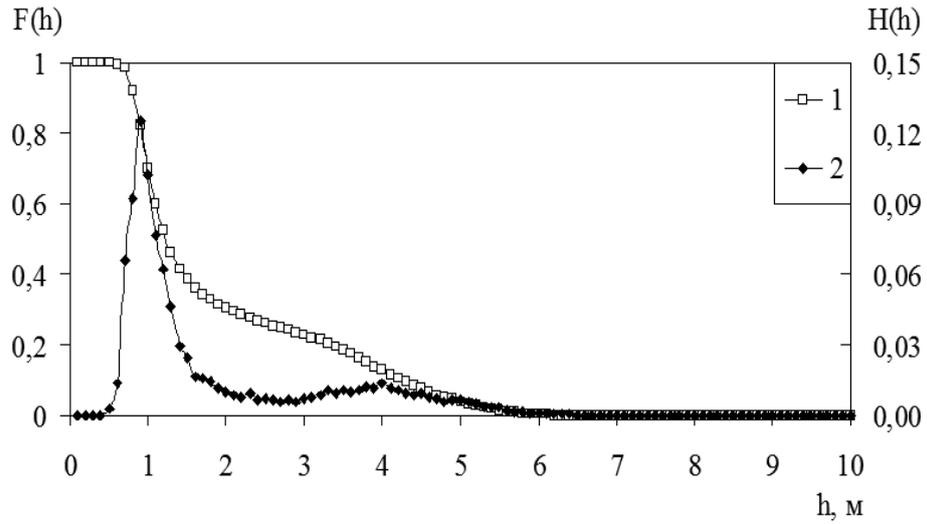

a)

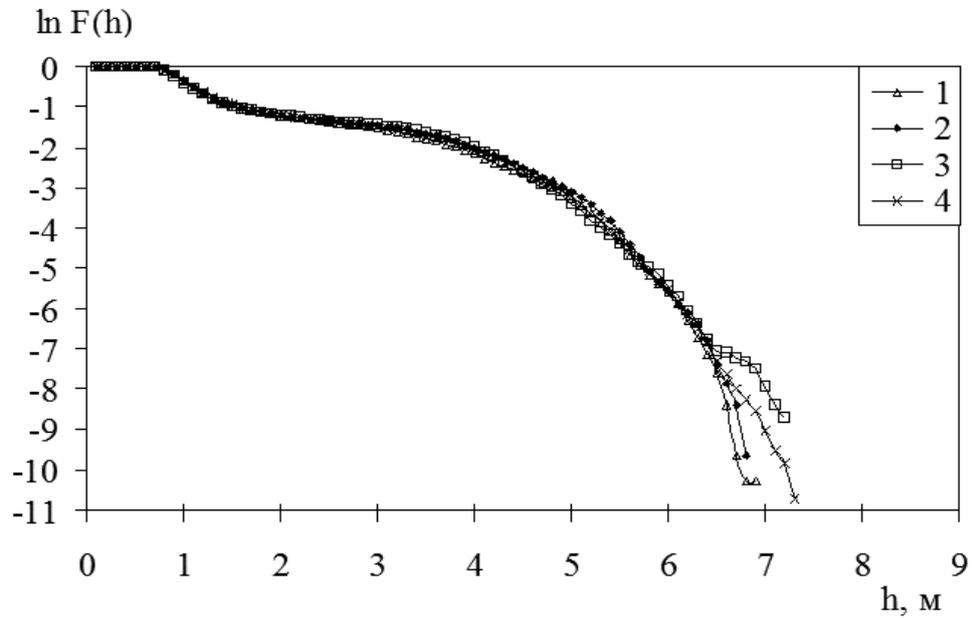

b)

Fig. 1. Statistics for the 30-year series of wind-wave heights at the central point of the Arabian Sea:
1a) Line 1 is provision function F; line 2 is histogram H;
1b) Part of the provision function F in the logarithmic coordinates. Lines 1, 2, 3 correspond to three different 10-years parts of certain initial 30-years series, and line 4 does for the whole 30-years series.

Taking the said into account, the general the requirement for selecting the sought domain of $F(W)$, suitable for constructing its approximation, can be determined by the condition

$$W_{lo} \leq W \leq W_{hi} \leq W_M .\qquad(2)$$

Here $W_{lo}$ and $W_{hi}$ are the lower and the upper edges of the proper domain of $F(W)$ used for constructing approximation $F_{ap}(W)$, which serve as the defining parameters of the latter. The domain [$W_{lo}$, $W_{hi}$], defined by ratio (2), hereafter is called as the domain of the provision function approximation. Its choice is to be specified, what will be discussed below.

In addition to the above, it is essential to note that the provision function can manifest a strong variability in the domain near the maximum value of the series $W_M$, while varying a length of the series. This feature of behavior for the provision function "tail» is clearly visible in the plot built in logarithmic coordinates for different time-series taken from the full one (Fig. 1b). This feature of $F(W)$ should also be taken into account when choosing the domain given by ratio (2).

Note also that for the reliability of approximation $F_{ap}(W)$, we need reliable $F(W)$. Therefore, to maximize the reliability of the "tail» for $F(W)$, the histogram should be built on a largest sample of data, i.e. on the full available series, rather than on a specially selected sample of peaks of the series, as it is used in a number of known methods (eg, AMS, POT and other common methods [2, 6, 7]).

Second, dwell on the question of choosing distributions. Traditionally [2, 7], analytical approximation $F_{ap}(W)$ is built by attracting any fixed known distribution function (Weibull, Pareto, log-normal, Gumbel, etc.). Herewith, as a rule, the justification of choosing the distribution used is given very arbitrary, do not taking into account specific local features of the process investigated [1, 7]. Exceptions are represented by very cumbersome techniques of the kind of BOULVAR [2,6].

Unknown parameters adopted in the distribution are determined by the least squares method, often called as the maximum likelihood method - MLM. This allows to minimize the error of approximation to some extent. However, the fixed feature of distribution does not permit to optimize the accuracy of approximation.

Finally, dwell on the reliability of return values estimates. In the literature [1-9], the procedure approximation optimization is never discussed, including selection of discreteness $\Delta W$ for histogram $H(W_i)$, quality and accuracy of the "tail" of which determines essentially the quality and accuracy of the analytical approximation, $F_{ap}(W)$. In this direction, further additional clarifications are quite needed.

Let us briefly summarize the main shortcomings of traditional approaches:
1) the weak validity of choosing the fixed statistics used for constructing approximations $F_{ap}(W)$ of provision function $F(W)$, performed regardless to its actual variability depending on the scales of quantities considered;
2) the lack of optimizing the accuracy of approximation $F_{ap}(W)$;
3) the lack of criteria ensuring reliability of approximation $F_{ap}(W)$ and its extrapolation.

The shortcomings named lead naturally to distortions and ambiguities in constructing analytical approximation $F_{ap}(W)$, that has an adverse effect on the accuracy of estimating return values of the studied series.

Farther we will offer a version of solution the issues mentioned, proposing the method which is more efficient in terms of simplicity and controlling accuracy and reliability of the return values estimates found for the series tested.

**3 The proposed method**

3.1 Choosing the provision function domain used for approximation

Taking into account the multiscale and unpredictable features of statistics for geophysical series, while constructing approximation for provision function $F(W)$, one should determine the parameters of ratio (2). Let us consider this issue on example of the histogram $H(H_S)$ for 30-year series of the significant wind-wave heights, $H_S$, given at the central point of the Arabian Sea. This $H(H_S)$, shown in Fig. 1a, has the second peak allocated at the point of wave height $H_S$ = 4.5m. Consequently, lower limit $W_{lo}$ of the ratio (2) cannot be less than 4.5m. More precisely this problem is solved by the following way.

Firstly, due to extreme smallness of magnitudes of the provision function "tail", it is worthwhile to construct an approximation of the "tail" in the logarithmic coordinates rather than approximation of the actual tail. This allows to construct approximation $F_{ap}(W)$ being more weighted for small values of $F(W)$. Moreover, in logarithmic coordinates, the "tail" of function $F(W)$ manifest plenty of twists (Fig. 1b). In our example, these twists appear at significant wave heights $H_S$ equal to values of 6.7 and 7.0 m, which are far beyond the second mode of the histogram but less the extreme value of the series, $H_{SM}$, being equal to 7,3m. This suggests that the wave-heights statistics vary on these scales, what should be accounted while choosing parameters of the approximation domain: $W_{lo}$, $W_{hi}$. Thus, on the example of given series, the most reasonable is to vary the quantities $W_{lo}$, $W_{hi}$ within the interval: $6,0 \leq W_{lo} \leq W_{hi} \leq 7,3$.

*3.2 Selection of a statistical distribution.*

In our mind, a more adequate way of constructing approximations $F_{ap}(W)$ is to put the logarithm of provision function in the form of a polynomial of degree $n$, the value of which may vary. Such variation the degree of polynomial, together with varying the domain (2) used for approximation constructing, allows the possibility to get an approximation with accuracy higher than in the case of using the known statistical distributions.

In this case, we restore the statistical distribution with the provision function of the form

$$F_{ap}(W) = exp\left[\sum_{i=0}^{i=n} a_i W^i\right] \quad . \tag{3}$$

Note that provision function of form (3) differs significantly from those known for common distributions. It can be considered as a kind of analog for the known Weibull distribution [2], or as a variation of the modified Weibull distribution having the form [9]:

$$F(W) = exp\left[-\left(\frac{W-b}{a}\right)^\beta + c\right], \tag{4}$$

where $a$, $b$, $c$ and $\beta$ are the distribution parameters.

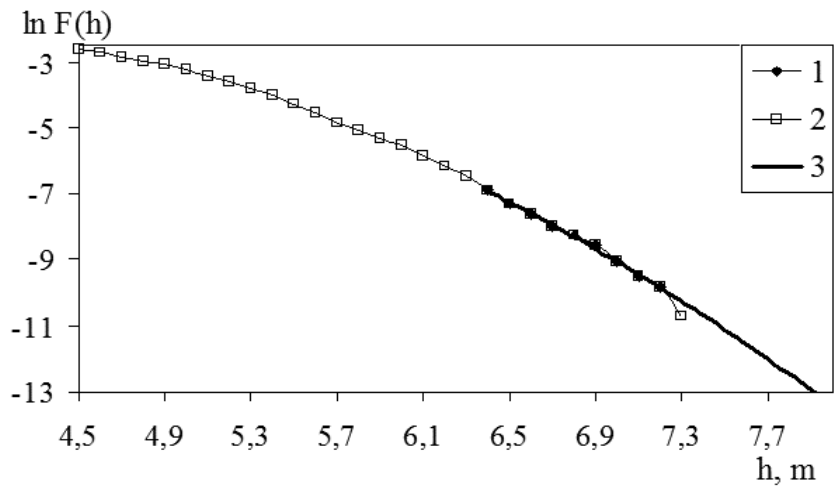

Fig. 2. An example of the polynomial approximation for the part of provision function F.
Line 1 is the domain of F, which is used for approximation; line 2 is the tail-part of function F; line 3 is the polynomial approximation with the subsequent extrapolation.
The formula of approximation is $y = -0{,}5455x^2 + 3{,}7882x - 8{,}8348$,
and parameter of statistical significance $R^2 = 0{,}996$.

An example of the proposed approximation, along with its subsequent extrapolation, is shown in Fig. 2. It is seen that the polynomial approximation follows rather well to the shape of "tail» for $F(W)$ at its chosen domain described by ratio (2) (the value of statistical significance $R^2$ differs from its limiting value, being equal to 1, in the third digit, only). It is clear that variation for parameters of domain (2) and degree of the polynomial $n$ allows controlling accuracy of the approximation $F_{ap}(W)$ and achieving its optimum value. This property of the proposed approach allows to optimize the process of constructing $F_{ap}(W)$.

The only disadvantage of the polynomial approximation $F_{ap}(W,n)$ is a necessity to control reliability of its extrapolation, as far as, in the case of considerable variability of the "tail» for $F(W)$, extrapolation of polynomial, having the order $n > 1$, can have twists and extremes. Such an extrapolation is implausible, of course. Therefore, it is necessary to vary the domain defined by relation (2) and the order of polynomial *n* in a manner avoiding the twists of extrapolation. The control mentioned can be easily performed both visually and by using the proper software.

*3.3 The quality of the analytical approximation*

Let us specify the concept of quality for approximation $F_{ap}(W)$, as far as such a concept is not mentioned in the literature [1, 2, 6-9]. We assume that the quality of approximation $F_{ap}(W)$ includes the following features.

1) The standard deviation $\delta$ (hereinafter, the deviation) which is determined by the magnitude of the differences between $F_{ap}(W)$ and $F(W)$ in logarithmic coordinates, given by the formula[1]:

$$\left( \frac{1}{N_T} \cdot \sum_{W_i=W_{lo}}^{W_i=W_{hi}} \left[ ln(F(W_i)) - ln(F_{ap}(W_i)) \right]^2 \right)^{1/2} = \delta, \qquad (5)$$

where

$$N_T = N_T(W_{hi}, W_{lo}) = (W_{hi} - W_{lo})/\Delta W \qquad (6)$$

is the number of points in domain (2) used for building approximation $F_{ap}(W)$. Obviously, the less $\delta$ the higher approximation accuracy, and consequently, the approximation is more preferable.

2) The condition of consistency given by the relation

$$F(W_M) \leq F_{ap}(W_M) \leq F(W_M) + \Delta, \qquad (7)$$

where $\Delta$ is the statistical uncertainty for the value of provision function at the last point, $F(W_M)$, defined by the maximum possible change of the histogram at the last point, $H(W_M)$, provided by a change of its discrete $\Delta W$.

3) The "plausibility" for extrapolation of $F_{ap}(W)$, implying the absence of twists and extrema on the extrapolation line for values $W > W_M$.

Let us give some explanations for the above criteria of approximation quality.

First. Deviation $\delta$ defined by formula (5), of course, depends on the choice of parameters for $F_{ap}(W)$: namely, the quantities of *n*, $N_T$ given by formula (6), and the "shift» $N_S$ defined by the formula

$$N_S = (W_M - W_{hi})/\Delta W \quad . \qquad (8)$$

---
[1] Note that estimate (5) could be modified with account of the statistical weight for provision function. This point is needed to be studied especially.

The optimal choice of the parameters mentioned allows easily to achieve minimization of $\delta$, providing the optimal estimates of return values, obtained by this method. All known methods [2, 6-9] have no such optimization.

Second. Consistency condition (7) means the requirement that the probability of occurrence the maximum value of the variable, $W = W_M$, given by the value of approximation $F_{ap}(W_M)$, must be upper the actual value of provision function, $F(W_M)$, existing the series. If condition (7) is not satisfied, it means the following:

a) Either, the value of approximation $F_{ap}(W_M)$ is less than the probability of occurrence the existing extreme value, $W = W_M$, what does not correspond to the reality; i.e. the approximation contradicts to the observation (this is the reason of using the term adopted);

b) Either, the approximation exceeds unnecessarily the really existing probability of extreme value $F(W_M)$; what is also considered as an unacceptable case.

Note that in criterion (7), the statistical uncertainty $\Delta$ for value $F(W_M)$ plays the key role. For its evaluation, one can use the following argument. The highest value of $\Delta$ corresponds to the case when the last interval of histogram $H(W_i)$ (i.e. the interval $[W_M \leq W \leq W_M + \Delta W]$) is represented by a single magnitude of a series, only, i.e. $H(W_M)=1/N$, where $N$ is the length of the series[2]. Then, it is natural to assume that the uncertainty of probability $F(W_M)$ (the latter is equal to $H(W_M)$), is determined by the possible realization of the random process for which the last interval of histogram $H(W_i)$ has two magnitudes, i.e. $H(W_M) = 2/N$. In this case, on the natural-logarithmic coordinates, the change of probability for the event $W=W_M$ (the sought uncertainty) is

$$\Delta = \ln(2/N) - \ln(1/N) = \ln(2) \approx 0{,}7 \qquad . \qquad (9)$$

It is this value is proposed to use in criterion (7), at this stage of the method development. In the future, this issue will be studied in more details.

Finally, we dwell on the subject of extrapolation plausibility. As already noted, the plausibility is an obvious requirement caused by the shortage of polynomial approximation, related to a possible appearance of twists or extrema in the course of constructing the extrapolation. This deficiency can be manifested only for values $n > 1$. Our experience of application the method shows that the appearance of twists and extrema in the extrapolation is usually accompanied either by a the deterioration of accuracy (leading to increasing deviation $\delta$) or by the violation of condition (7). The meeting both requirement: minimizing $\delta$ and condition (7), as a rule, allows a rejection approximations $F_{ap}(W)$ leading to the "implausible" extrapolation. In automatic mode, this problem is solved by using proper software, i.e. by the numerical filtering, which will be implemented in future research.

---

[2] Otherwise, one can reduce the discrete of histogram, $\Delta W$, to the case needed. When coming into the last interval of the histogram two (or more) magnitudes, the value of uncertainty is reduced: $\Delta = \ln(3/2) \approx 0{,}4$ (or less); that allows us do not consider such cases.

Important to note that the implicit sign to a possible appearance the "implausibility" for extrapolation of polynomial approximation is an existence of zero values of histogram $H(W_i)$ in the range of values $W_i$ which are smaller than $W_M$ (i.e. the "internal zeros", as far as at the end of histogram, the nonzero value, $H(W_M)$, exists). Such a kind histograms are often obtained, when a large scattering of high values of variable $W$ takes place. It means that, in the case, too small value for discreteness of histogram $\Delta W$ is chosen. In this case, the presence of zeros in histogram leads to the fact that the "tail" of provision function $F(W)$ has a "shelf" (areas with a zero derivative). Hitting the "shelves" to the boundary of domain (2) can cause the "implausibility" for extrapolation of polynomial approximation, for a certain choice of parameters of approximation.

To avoid the situations said, it is necessary so to choose the discrete of histogram, $\Delta W$, in such a way that the mentioned defect of histogram $H(W_i)$, and consequently, one of the provision function $F(W)$, does not appear. Basing on the said, we can formulate the following rule for choosing value of histogram discrete $\Delta W$: the value of $\Delta W$ should be much smaller than the maximum value of variable $W_M$ (for example, more than 10 times smaller, to provide a reliable provision function by the formula (1) ), but it should be more than the smallest value $\Delta W_{mi0}$ which still ensures the absence of "internal" zeros on the "tail" of histogram $H(W_i)$. Herewith, if the selected value $\Delta W$ is less the accuracy of measurements for $W$, denoted as $\delta W$, the value of $\Delta W$ should be taken equal to $\delta W$. This condition can be formalized by the formula

$$\text{Max}\{\delta_W, \Delta W_{mi0}\} \leq \Delta W \ll W_M \quad . \tag{10}$$

However, the consistency condition (7) must be always satisfied.

### 4 Examples of application of the method

Let us present the results of applying the proposed method of estimating the regime values for wind speed $W$ and significant heights of waves $H$s, as well as the comparison of our results with those obtained by a known method of initial distribution [1, 6]. To do this, we attract the series of wind speed and wave heights specified in Section 2. The parameters of approximations and the minimal deviation parameters $\delta$ are given in Tab. 1, and the estimates of $W_M$ and return values are done in Tab. 2.

A detailed analysis of constructing approximations optimized with $\delta$ shows that the choice of all the approximation parameters ($n$, $N_T$ and $N_S$) is equally important. Herewith, the approximations constructed for low values of shear $N_S$ and values $N_T \gg 1$, used for the domain (2), have evident advantage in terms of accuracy of approximation δ. It is interesting to note that the varying a degree of the polynomial approximation does not strongly impact on the return values (for example, ones for 100 years). They are retained in the limits which are smaller the precision of measurements for the values (which is of the order of $δ_W ≈ 0.2$ m for the wave heights), and the histograms discrete, $ΔW$ (Fig. 3a, b). However, this does not negate the appropriateness of variation $n$ for the aim of optimizing parameter δ.

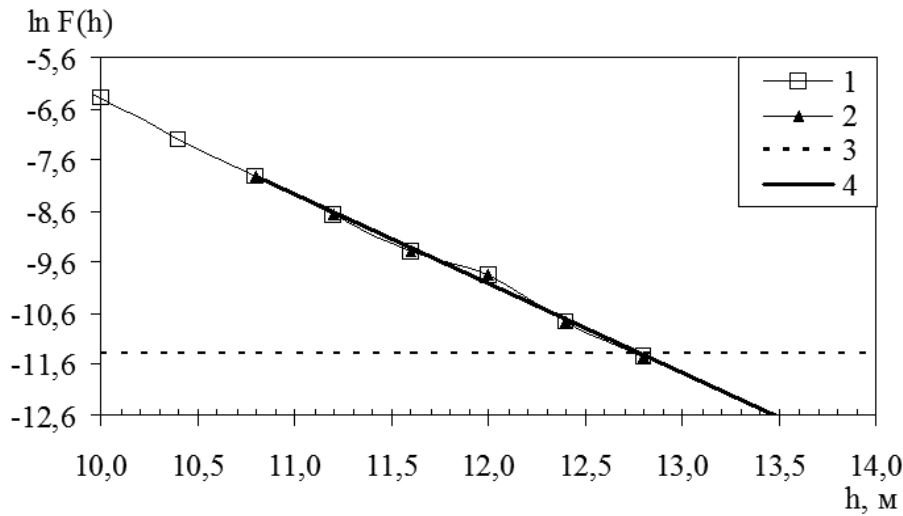

a)

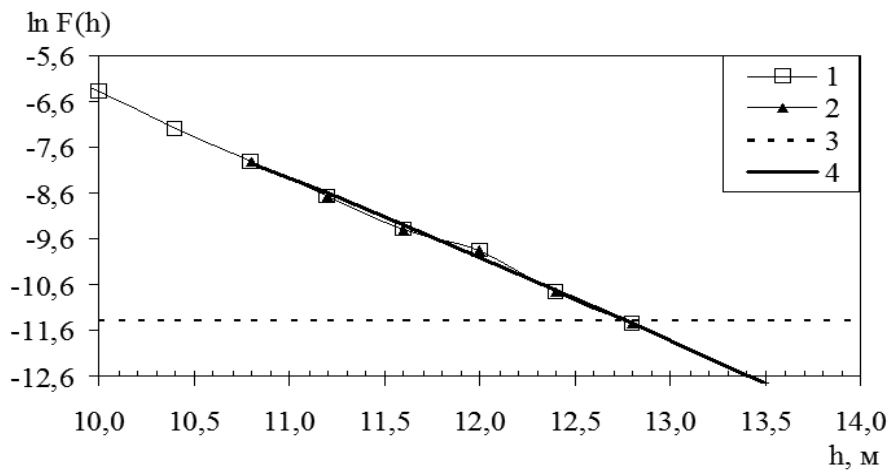

b)

Fig. 3. Illustration of differences in approximations for provision function F obtained for the central point in the zone 6 of the Indian Ocean (the main parameters of the approximations are: $N_T = 6$ $N_S = 0$) .
3a) The polynomial of the first order with features: y = -1,7489x + 10,959 and $R^2$ =0,9953;
3b) The polynomial of the second order with features: y = -0,0496$x^2$ -9,5781x +4, 0742 and $R^2$ =0,9955.
Line 1 is function F; line 2 is the domain of F, which is used for approximation:
line 3 shows the level of probability for the wave height returning once for 30 years (the horizontal axis corresponds to the level of wave height returning once for 100 years); line 4 is the polynomial approximation and its extrapolation to the horizontal axis.

Comparison of our estimates of regime values with those obtained for the same series by the method of initial Weibull distribution, used in [1], shows a noticeable difference[3]. Herewith, the regime values found by the known method, corresponding to 1 for30 years, are usually lower than the actual absolute maxima presenting for the considered 30-year period (see, Tab. 2). This allows to suggests that the method of initial Weibull distribution does contradict to the actual occurrence of extreme values. In the proposed method, this disadvantage is impossible principally, in view of the condition (7).

**5 Conclusions**

The proposed approach to the problem of estimating regime values for wind and wave series is based on the use polynomial approximation of provision functions of security, including variation parameters for this approximation, aimed to its optimizing based on a number of criteria formulated. This approach provides a new class of statistics with the provision function of form (3), allowing analytical extrapolating the "tails" of them.

In the work, the basic procedures of the proposed method are discussed, and the practical guidance is given, allowing its wide implementation (Section 3). However, some issues concerning the accuracy of the estimates (selecting the histograms discrete, evaluating accuracy of the provision function and its extrapolation, and others) do require a more thorough investigation.

Note that the listed unsolved issues said are equally related to both the proposed method and to all known methods of estimating the return values of the random geophysical series.

**Acknowledgments**

The authors thank the members of team of the RFBR grant № 14-05-92692-IND_a: Pogarsky FA, Saprykina Y., Kuznetsov SY for their assistance in the work and discussing the results. Separately, we are grateful to VV Fomin for provision the estimates obtained by the initial Weibull distribution. This work was supported by the RFBR project mentioned.

---

[3] The results of estimates for initial Weibull distribution are kindly provided to us by V. Fomin